\documentclass[twocolumn,twoside,slac_two]{revtex4}
\usepackage{graphicx}
\usepackage{fancyhdr}
\def\fermilat{\textit{Fermi}/LAT}

\def\gama{$\gamma$-ray }

\begin{document}

\title{The connection between the gamma-ray emission and millimeter flares in Fermi/LAT blazars}

%

\author{J. Le\'on-Tavares, M. Tornikoski, A. L\"ahteenm\"aki}
\affiliation{Aalto University Mets\"ahovi Radio Observatory,  Mets\"ahovintie 114, FIN-02540
    Kylm\"al\"a, Finland.}
\author{E. Valtaoja}
\affiliation{ Tuorla Observatory, Department  of Physics and Astronomy, University of Turku, 20100 Turku, Finland.}

\author{E. Nieppola}
\affiliation{ Finnish Centre for Astronomy with ESO (FINCA), University of Turku, V\"ais\"al\"antie 20, FI-Piikki\"o, Finland.}

\begin{abstract}
We compare the $\gamma$-ray photon flux  variability of northern blazars in the \fermilat\ First Source Catalog with 37 GHz radio flux density curves from the Mets\"ahovi quasar  monitoring program. We find that the relationship between simultaneous  millimeter (mm) flux density and $\gamma$-ray  photon flux is different for different types of blazars. The flux  relation between the two bands is positively correlated for quasars and does no exist for BLLacs. Furthermore, we find that the levels of $\gamma$-ray emission in high states depend on the phase of the high frequency radio flare, with the brightest $\gamma$-ray events coinciding with the initial stages of a mm flare. The mean observed delay from the beginning  of a mm flare to the  peak of the $\gamma$-ray emission  is about 70 days, which places the average location of the $\gamma$-ray production at or downstream of the radio core. We discuss  alternative scenarios for the production of $\gamma$-rays at distances of parsecs along the length of the jet . 

\end{abstract}

\maketitle

\thispagestyle{fancy}


\section{Introduction}
 Several studies, based on the first year of  Fermi/LAT  operations,   have shown that:  (i)   the $\gamma$-ray  and the averaged radio flux densities  are significantly correlated  \cite[e.g.][]{fermi_correlation} and (ii) blazars detected at $\gamma$-rays are more likely to have larger Doppler factors and larger apparent opening angles than those not detected by LAT \cite[e.g.][]{lister_2011}.  This observational evidence strongly suggests  that  radio and $\gamma$-ray  emission  have a co-spatial origin. To locate and identify the region where  the bulk of $\gamma$-ray emission is produced, and to provide details about  its connection to  the radio jet,  an analysis of  simultaneous radio and $\gamma$-ray light curves is essential. 

However, we highlight two caveats about the interpretation of radio/gamma correlation analyses, which have often been interpreted too simplistically. First, it is  well-known  that there is usually a considerable delay between mm and cm radio flares. Thus, although cm-flares would tend to peak after the $\gamma$-ray flares, mm-flares would show shorter delays or possibly even peak before the $\gamma$-rays. The very important second caveat is that a correlation analysis tends to measure the distance between the peaks, especially if the flares have different timescales (as the radio and the $\gamma$-ray flares tend to have ). However, a radio flare starts to grow a considerable time before it peaks. The \emph{beginning} of a millimeter flare  coincides with the ejection of a new VLBI component from the radio core  \cite{savolainen_2002}. This is the epoch that must be compared with the $\gamma$-ray flaring, not the epoch of the radio flare maximum. The crucial question is whether a $\gamma$-ray flare occurs \emph{before} the beginning of a mm-flare, or \emph{after} it; in the former case the $\gamma$-rays originate upstream of the radio core (the beginning of the radio jet), in the latter, they originate downstream of the radio core, presumably from the same disturbances that produce the radio outbursts. 

In this proceeding we summarize our recent results  \cite{mmflares}, obtained after combining  the finely  sampled  37~GHz Mets\"ahovi  light curves and the monthly binned $\gamma$-ray light curves provided by  the \fermilat\ First Source Catalog \cite[1FGL,][]{1FGL}. By using a radio flare decomposition method, we estimate the beginning epochs of millimeter flares   and their phases during $\gamma$-ray flaring events  to establish the true temporal sequence between $\gamma$-ray and radio flaring.

\begin{figure*}[!t]
\centering
\includegraphics[width=\textwidth]{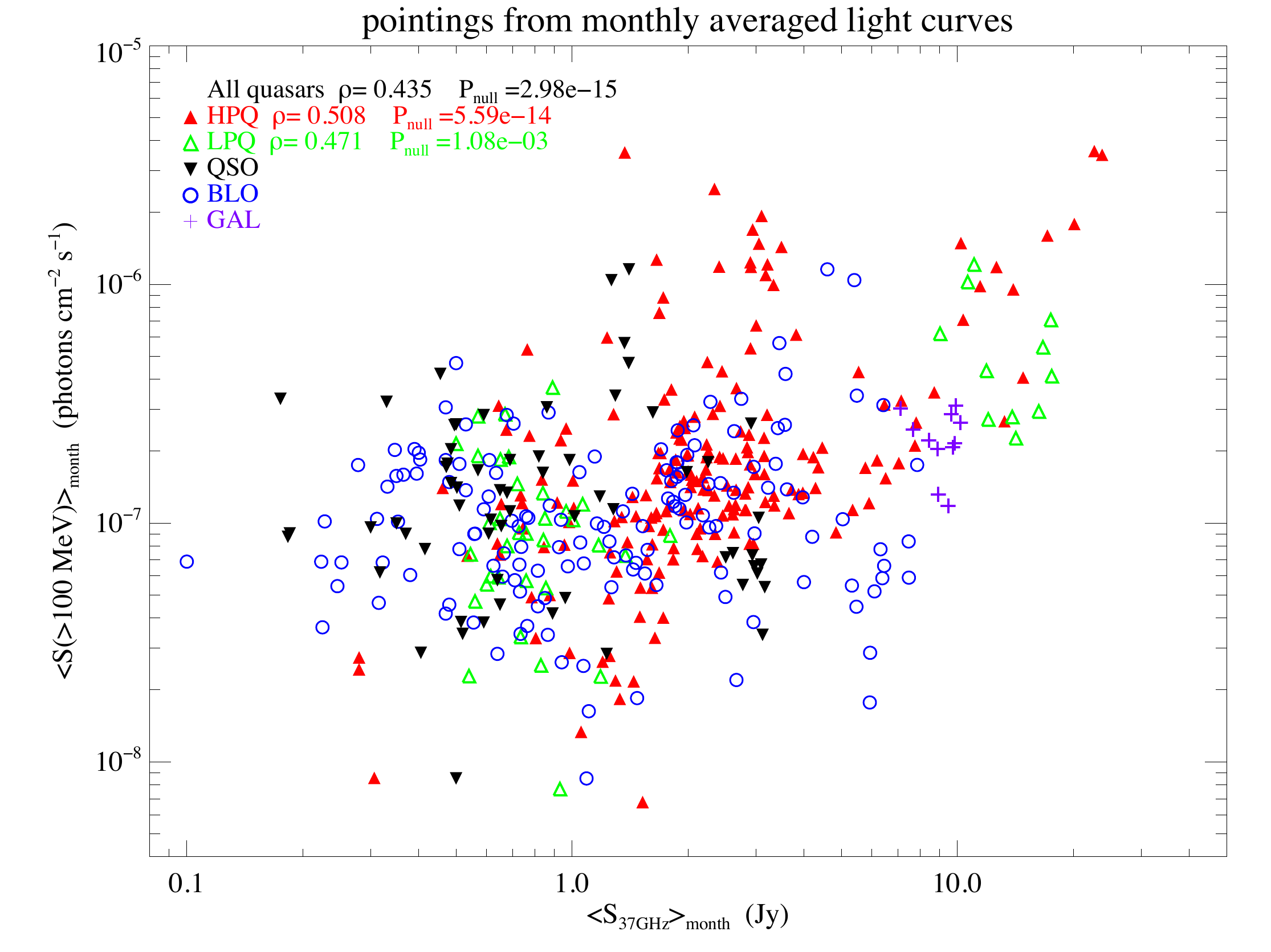}
\caption{Monthly flux-flux relation for the combined sample of 60  northern AGN. The different types of sources are symbol coded as shown in the legend, the correlations coefficients are shown only when the significance is $> 99.9 \%$.} \label{JACpic2-f2}
\end{figure*}

\section{The 37~GHz and \gama light curves}

The Mets\"ahovi quasar monitoring program  currently includes about 250 AGN at 37~GHz.  From them, we selected a sample of sources that fulfill the following criteria: (i)  well-sampled light curves during the period 2007-2010, covering the 1FGL period, (ii)  a firm association with the 1FGL catalogue, and (iii) a $\gamma$-ray monthly light curve during the 1FGL period that is significantly different from a flat one.

Our final sample consists of 60 sources  classified according to their optical spectral type as highly polarized quasars (HPQ, 22), low polarization quasars (LPQ,~5), quasars without any  optical polarization data (QSO,~15), BL~Lac type objects (BLO,~17), and radio galaxies (GAL,~1).  The sample of sources used in this work   is listed in Table 1 of \cite{mmflares}

\section{The S$_{37GHz}$ - S$_{\gamma}$ correlation}

To compare radio flux densities  and $\gamma$-ray photon fluxes, monthly binned radio light curves were created from the Mets\"ahovi light curves. The time bins are the same as in the 1FGL flux  history curves, allowing us to compare simultaneously $\gamma$-ray photon flux  and radio flux-density variations with the time resolution of a month.

 \begin{figure*}[!t]
\centering
\includegraphics[width=\textwidth]{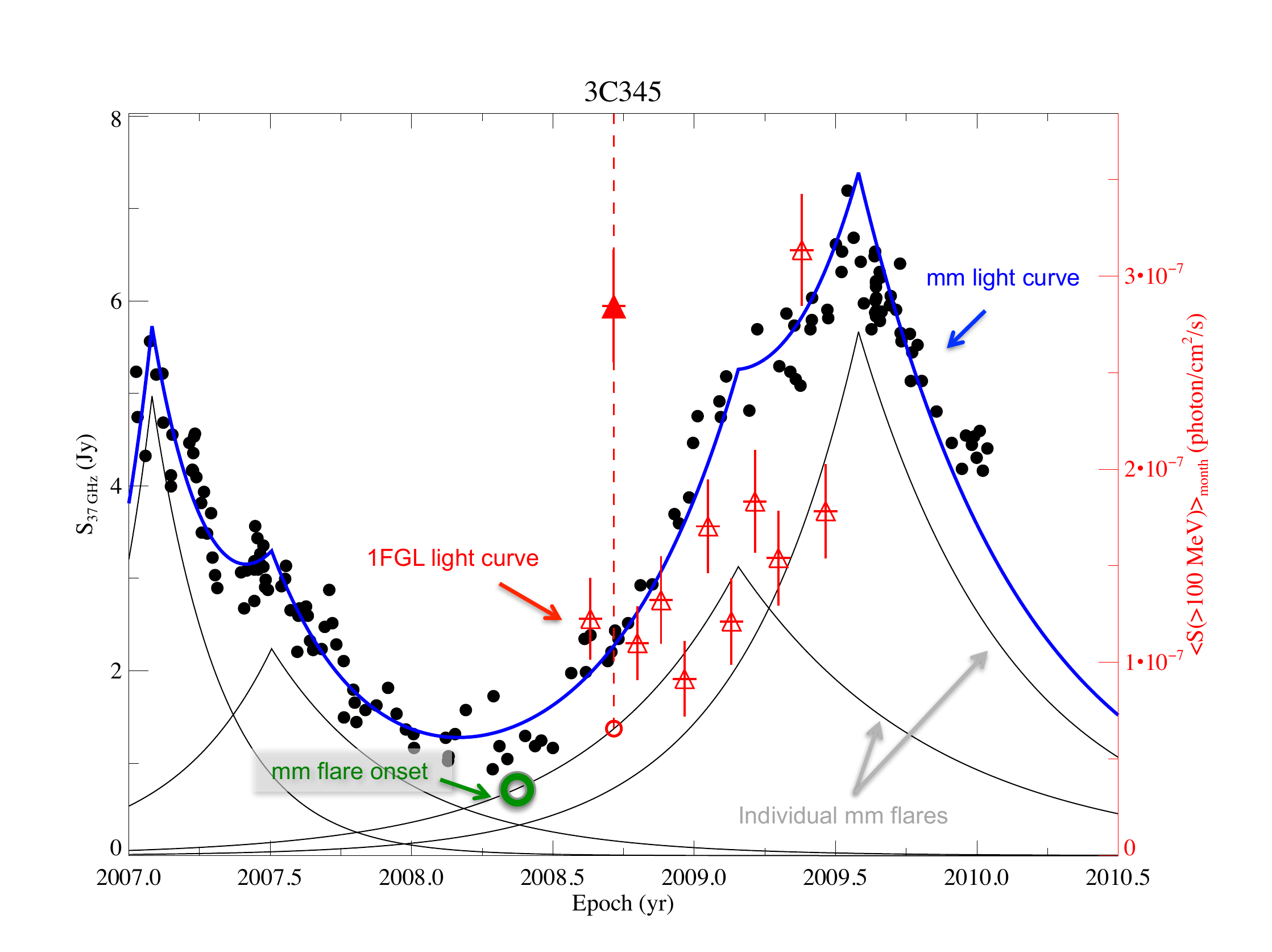}
\caption{The recent flux  history at 37~GHz (filled circles) and  $\gamma$-rays (triangles) of 3C~345.  The total flux density  has been decomposed using individual exponential flares. The filled triangle represents the most significant peak in the $\gamma$-ray flux density during the 1FGL period. The dashed vertical line is drawn to highlight the relation between the peak in the $\gamma$-ray light curve and the ongoing millimeter flare.} 
\end{figure*}

 Figure 1 shows that simultaneous measurements at the two bands appear to be positively correlated. However, we find significant differences between quasars and BL~Lacs, which we describe below.  By applying the Spearman's rank correlation test, two very clear results emerge. First, there is a significant positive correlation between the $\gamma$-ray  photon flux and the 37~GHz flux density for quasars, while the BL~Lac fluxes  are not correlated. Second, the strength and the significance of the correlations is different for each type of quasar.

The photon flux - flux density  correlation for quasars is absent for QSOs, significant  for LPQs, and very significant for HPQs. Such a dependence on the degree of optical polarization may arise naturally  if the polarization indicates the viewing angle of the jet, with sources with high optical polarization having their jets oriented closest to our line of sight (see Figure 7 in \cite{elina_2011}). The dependence of the flux - flux relation on optical polarization agrees with previous results, where it has been shown that the brightest $\gamma$-ray emitters have preferentially smaller viewing angles   and consequently higher Doppler boosting factors \cite[e.g.,][]{lister_2011}. Since $\gamma$-ray fluxes and radio flux densities are significantly correlated for sources where the relativistic jet is aligned close to our line of sight, this implies that there is  a strong coupling between the radio and the $\gamma$-ray emission mechanisms. That the correlation is seen on monthly timescales further indicates a cospatial origin in quasar-type blazars.

\section{Connections between ongoing flares and high states of
  gamma-ray emission}

We have decomposed the mm light curves into individual  exponential flares (see Figure 2), each of which corresponds to a new disturbance created in the jet and is often detectable as a new VLBI component \cite{savolainen_2002}. We further calculate the phase of the mm-flare when  the most prominent maxima in the 1FGL light curves occurred. Our analysis shows  that the most significant $\gamma$-ray flux peaks tend to occur when a mm-flare is either rising or peaking.  This indicates that the strong $\gamma$-ray flares are produced in the same disturbances that produce the mm flares.

 Using our data, we can estimate the time delay between the time when the mm flare starts and when the $\gamma$-rays peak for each source. We define the beginning of a mm flare to be:
\begin{equation}
t_{0}^{mm} = t_{max}^{mm} - \tau,
\end{equation}
where  $t_{max}^{mm}$ is the time of the mm flare peak and $\tau$ is the variability timescale. In other words, we define the beginning of the flare as the epoch when its flux is $1/e$ of the maximum flux, $S(t_{0}^{mm}) = \frac{S_{max}^{mm}} {e}$.  For each source we estimate the time delay between    the time of  mm-flare onset (green circle in Figure 2)  and when the $\gamma$-ray peak occurs (filled triangle in Figure 2). The observed time delay has a distribution centered around 70 days with the onset of the  mm-flare  preceding the $\gamma$-ray peak.

After converting the time delays to linear distances from the region where the mm-outburst begins (i.e. the radio-core) to the region of the $\gamma$-ray production, our  estimates lead us to conclude that in our sample  the average location of the $\gamma$-ray emission region is  about 7 parsecs downstream the radio-core. 

\bigskip

\section{Summary and Discussion}

The results presented in \cite{mmflares} strongly indicate that at least for the strongest $\gamma$-rays the production sites are downstream or within the  radio core \cite{pushkarev_2010}, well outside the BLR at distances of several parsecs or even tens of parsecs from the black hole and the accretion disk. A number of papers based on  \emph{Fermi} data have reached similar conclusions \cite[e.g.][]{oj287}.

In the current AGN paradigm, it is widely believed that \gama are produced via Inverse Compton (IC) mechanisms in the relativistic  jet,  and  the  seed photons  for the IC process may be provided    by the jet synchrotron emission (Synchrotron Self Compton, SSC) as well as by external photon sources (External Compton, EC) such as an accretion disk, the  broad-line region (BLR)  or the  hot dusty torus. At the distance of 7 parsecs from the radio core --  keeping in mind that the radio core itself is often at a considerable distance from the black hole \cite[e.g.][]{marscher_2008} --  the only sources of seed photons for the IC processes are the jet itself and the dusty torus.   However,  there is growing evidence  \cite{3c120, 3c3903}  that  the BLR might extend to much larger distances than given by virial estimates ($R_{BLR} < 1~pc$). One possibility is that  the jet   drags a part of the BLR whit it, see Figure 3 for a sketch  of  an outflowing BLR among the other inner AGN constituents.

\begin{figure}[t!]
\centering
\includegraphics[width=\columnwidth]{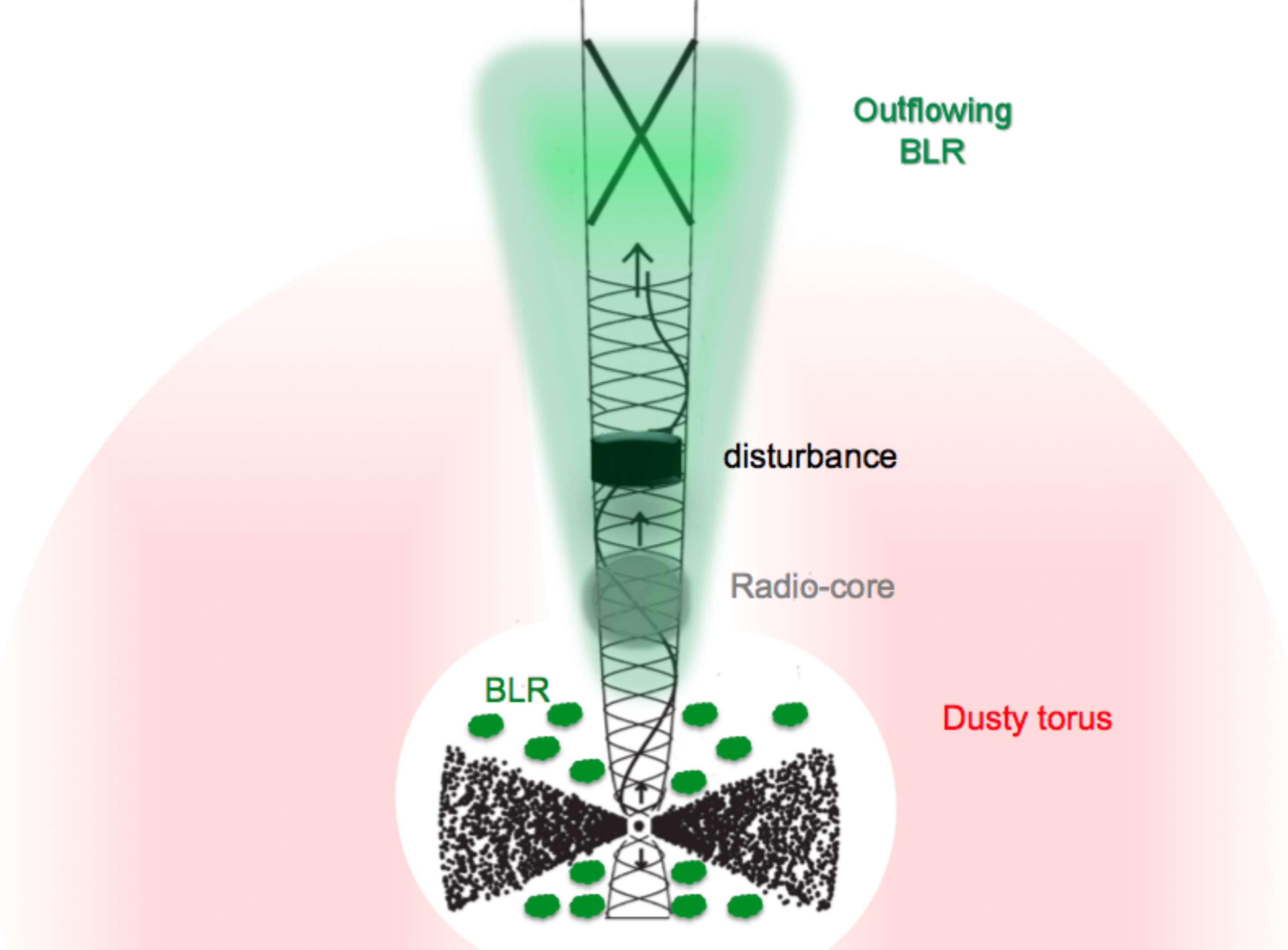}
\caption{Sketch of the inner parts of an AGN. Recent results  \cite{3c3903,3c120} have shown that a significant amount of broad-line emission is driven by  optical continuum radiation  produced downstream of the radio core, this strongly suggests the presence of  an additional component of the BLR located at parsec scales. In this scenario, an outflowing BLR may serve as a source of seed-photons to produce $\gamma$-rays via EC models  far away from the central engine \cite{mmflares}.} 
\end{figure}

While single zone  SSC has failed to reproduce the observed $\gamma$-rays \cite{lindfors_2005} and so far there are only a couple of blazars with firm detections of  a dusty torus \cite{marc_2006,malmrose_2011}, a tentative  idea to test  is  \emph{ whether an  outflowing BLR  can serve  as a source of external photons to produce $\gamma$-rays, even at distances of parsecs downstream of the radio-core}. In this scenario, the strong $\gamma$-ray events are produced in the  same disturbance that produces the radio outburst   by upscattering  external photons  provided by an outflowing BLR.

The  most effective way to explore any of the above  scenarios (and others) is  by modeling  simultaneous, well-sampled   spectral energy distributions (SEDs) \cite{planck_sed,giommi_2011} following  a multizone modeling approach   \cite{marc_2011}.   Furthermore,  recent results   have suggested  that the more massive the black hole is, the faster and the more luminous jet it produces \cite{bllacs}. Therefore,  a reliable estimate of  the black hole masses  is an essential input  to theoretical models of both the shape and the variability of blazars SEDs.

\bigskip
\begin{acknowledgements}
We acknowledge the support from the Academy of Finland to our AGN monitoring project (project numbers 212656, 210338, 122352 and others).   
\end{acknowledgements}


\bigskip
 

\begin{thebibliography}{18}   
\bibitem{fermi_correlation}Ackermann, M., Ajello, M., Allafort, A., et al., {arXiv:1108.0501}  (2011) 
\bibitem{lister_2011} Lister, M.~L., Aller, M., Aller, H., et al., {arXiv:1107.4977} (2011)
\bibitem{savolainen_2002} Savolainen, T., Wiik, K., Valtaoja, E., Jorstad, S.~G., \& Marscher, A.~P., \emph{A\&A}, 394, 851 (2002) 
\bibitem{mmflares} Le{\'o}n-Tavares, J., Valtaoja, E., Tornikoski, M. et al., \emph{A\&A}, {532}, A146 (2011)
\bibitem{1FGL} Abdo, A.~A., Ackermann, M., Ajello, M., et al., {ApJS}, 188, 405 (2010) 
\bibitem{elina_2011} Nieppola, E., Valtaoja, E., Tornikoski, M. et al.,  { arXiv:1109.5844} (2011)
\bibitem{pushkarev_2010} Pushkarev, A.~B., Kovalev, Y.~Y., \& Lister, M.~L., {ApJL}, 722, L7 (2010) 
\bibitem{oj287} Agudo, I., Jorstad, S.~G., Marscher, A.~P., et al., \emph{ApJL}, 726, L13 (2011)
\bibitem{marscher_2008} Marscher, A.~P., Jorstad, S.~G., D'Arcangelo, F.~D., et al., {Nature}, 452, 966 (2008)
\bibitem{lindfors_2005} Lindfors, E.~J., Valtaoja, E., {\ T\"u}rler, M., {A\&A}, 440, 845 (2005) 
\bibitem {marc_2006} T{\"u}rler, M., Chernyakova, M., Courvoisier, T.~J.-L., et al., {A\&A}, 451, L1 (2006)
\bibitem{malmrose_2011} Malmrose, M.~P.,  Marscher, A.~P., Jorstad, S.~G., Nikutta, R.,  \& Elitzur, M., {ApJ}, 732, 116 (2011)
\bibitem{3c3903} Arshakian, T.~G., Le{\'o}n-Tavares, J., Lobanov, A.~P., et al., \emph{MNRAS}, 401, 1231  (2010)
\bibitem{3c120} Le{\'o}n-Tavares, J., Lobanov, A.~P., Chavushyan, V.~H., et al., \emph{ApJ}, {715}, 355 (2010) 
\bibitem{planck_sed} Planck Collaboration, Aatrokoski, J., Ade, P.~A.~R., et al., {arXiv:1101.2047 },, (2011) 
\bibitem{giommi_2011} Giommi, P., Polenta, G.,  Lahteenmaki, A., et al., {arXiv:1108.1114},  (2011)
\bibitem{marc_2011}T{\"u}rler, M., \& Bj{\"o}rnsson, C.-I., {arXiv:1109.2518}, (2011) 
\bibitem{bllacs}  Le{\'o}n-Tavares, J., Valtaoja, E., Chavushyan, V.~H., et al., \emph{MNRAS}, 411, 1127 (2011)



\end{thebibliography}
\end{document}